\documentclass[a4paper, amsfonts, amssymb, amsmath, reprint, showkeys, nofootinbib, twoside,superscriptaddress,prx,citeautoscript,longbibliography]{revtex4-2}
\usepackage[pdftex, pdftitle={Article}, pdfauthor={Author}]{hyperref}

\usepackage{graphicx}
\usepackage{bm}
\usepackage{xcolor}

\begin{document}

\preprint{APS/123-QED}

\title{Entropy as a Design Principle in the Photosystem II Supercomplex}
\author{
    Johanna L. Hall$^{a,b,c}$, Shiun-Jr Yang$^{a,b,c,f}$, David T. Limmer$^{a,c,d,e}$, \\
    and Graham R. Fleming$^{a,b,c,}$\thanks{Corresponding Author: Graham R. Fleming, Hildebrand Hall 221, University of California, Berkeley, Berkeley, CA 94720, (510) 520-4220, \texttt{grfleming@lbl.gov}}
}

\affiliation{$^a$Department of Chemistry, University of California, Berkeley, Berkeley, CA 94720. \\
$^b$Molecular Biophysics and Integrated Bioimaging Division, Lawrence Berkeley National Laboratory, Berkeley, CA 94720. \\
$^c$Kavli Energy Nanoscience Institute at Berkeley, Berkeley, CA 94720. \\
$^d$Chemical Sciences Division, Lawrence Berkeley National Laboratory, Berkeley, CA 94720. \\
$^e$Materials Sciences Division, Lawrence Berkeley National Laboratory, Berkeley, CA 94720. \\
$^f$Current Affiliation: Department of Chemistry, Massachusetts Institute of Technology, Cambridge, MA 02139.}

\vspace{1em}

\date{\today}

\begin{abstract}
Photosystem II (PSII) can achieve near-unity quantum efficiency of light harvesting in ideal conditions and can dissipate excess light energy as heat to prevent formation of reactive oxygen species under light stress. Understanding how this pigment-protein complex accomplishes these opposing goals is a topic of great interest that has so far been explored primarily through the lens of the system energetics. Despite PSII's known flat energy landscape, a thorough consideration of the entropic effects on energy transfer in PSII is lacking. In this work, we aim to discern the free energetic design principles underlying the PSII energy transfer network. To accomplish this goal, we employ a structure-based rate matrix and compute the free energy terms in time following a specific initial excitation to discern how entropy and enthalpy drive ensemble system dynamics. We find that the interplay between the entropy and enthalpy components differs among each protein subunit, which allows each subunit to fulfill a unique role in the energy transfer network. This individuality ensures PSII can accomplish efficient energy trapping in the RC, effective NPQ in the periphery, and robust energy trapping in the other-monomer RC if the same-monomer RC is closed. We also show that entropy, in particular, is a dynamically tunable feature of the PSII free energy landscape accomplished through regulation of LHCII binding. These findings help rationalize natural photosynthesis and provide design principles for novel, more efficient solar energy harvesting technologies.
\end{abstract}

\maketitle

\section{Introduction}
Photosystem II of land plants and other oxygenic photosynthetic organisms possesses the unique ability to split water and generate molecular oxygen. The reaction center (RC), where charge separation is initiated, is coupled to an antenna containing several hundred chlorophyll (Chl) molecules to optimize photosynthetic yield in varying light levels. This pigment-protein supercomplex can convert harvested light energy into a charge separation with near perfect quantum efficiency\cite{1-blankenship2022molecular,2-wientjes2013quantum,3-vanmieghem1995charge}. In light stress conditions, Photosystem II (PSII) is able to dissipate excess light energy as heat to prevent the formation of reactive oxygen species (ROS) via reaction of molecular oxygen with the chlorophyll triplet excited state\cite{4-muller2001nonphotochemical,5-durrant1990characterisation,6-krieger2004singlet}. Many of the photoprotective capabilities are hypothesized to take place in the periphery of PSII because of the high oxidative potential of the RC components\cite{7-ruban2016nonphotochemical,8-ishikita2005redox,9-genty1989relationship,10-ruban2021mechanism}, which limits the ability to place photoprotective molecules close to the RC. In this study, we explore how entropic driving forces facilitate these opposing design requirements of PSII.

Perhaps the most intuitive design for an antenna/reaction center system is that of an energy funnel with the reaction center located at the energy minimum. Indeed this is precisely the approach adapted by the purple bacterial B800/B850/RC system, for example\cite{1-blankenship2022molecular,11-mirkovic2017light}. PSII, however, has a multi-subunit construction with the lowest energy levels of all the subunits having very similar energies1 (Figure S1). This has led to descriptions of the PSII antenna system as a flat landscape\cite{12-yang2024design,13-leonardo2024bidirectional}, as a very shallow funnel\cite{14-jennings1993study,15-jennings1993distribution}, or as lacking a funnel\cite{16-bennett2013structure,17-shibata2013photosystem,18-croce2014natural}. This, in turn, raises the question of the driving force for energy transfer to the RC and the potential role of entropic in addition to enthalpic contributions. The availability of high-resolution structures for the 204 Chl-containing C$_2$S$_2$-type and 312 Chl C$_2$S$_2$M$_2$-type dimeric supercomplexes\cite{19-wei2016structure,20-su2017structure}, along with quantitative models for the excitation dynamics for both structures\cite{12-yang2024design,13-leonardo2024bidirectional}, allows us to explore the free energy landscape of the supercomplexes. Here, we use these dynamical models to study the stochastic thermodynamics accompanying photorelaxation and quantify the time-dependent changes to both the energy and entropy. These ensemble-level calculations give insight into the evolution of the initial exciton and we find that, qualitatively, the ensemble evolution can be thought of in two phases. First, an entropic exploration of the landscape distributes population amongst the chromophores with similar energies. Second, a more directed motion takes population down the shallow energy gradient toward the reaction center. We also show that when one of the two RCs is closed through prior excitation, the entropic phase on the side of the dimeric structure with a closed RC is elongated, allowing more time for the excitation to find the second, open, reaction center. We also show how the entropy component of the free energy landscape is a tunable parameter, which PSII can alter by regulating the binding of peripheral antenna complexes\cite{21-ballottari2012evolution,22-boekema1999multiple,23-boekema1999supramolecular,24-caffarri2009functional,25-crepin2018functions,26-dekker2005supramolecular}. These calculations provide insight into how the design of the PSII supercomplex enables the combination of both efficient charge separation and effective photoprotection, and how these design principles can be applied to the construction and characterization of general energy harvesting systems. 

\begin{figure}
    \centering
    \includegraphics[width=8.5cm]{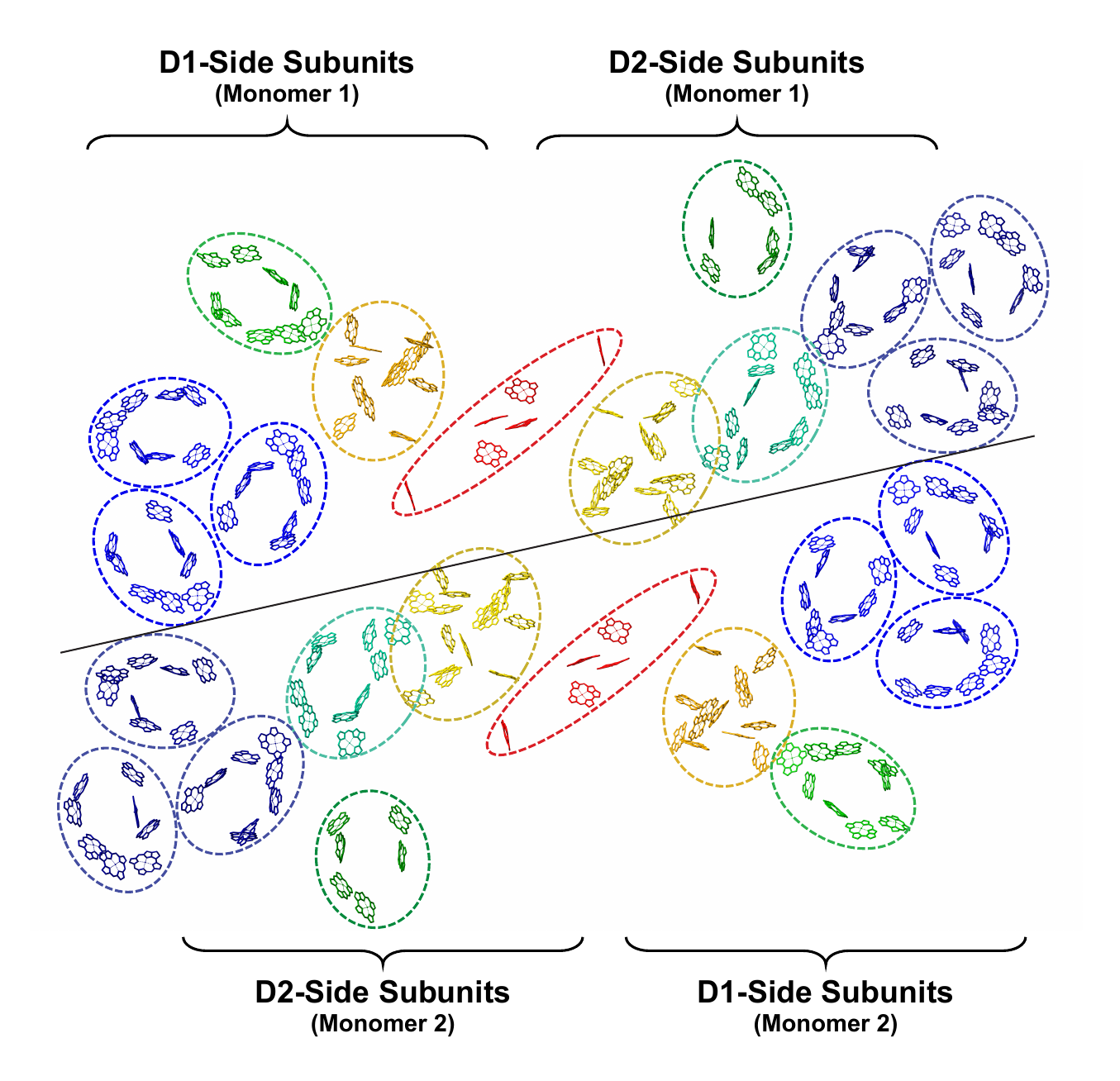}
    \caption{The pigment arrangement of the C$_2$S$_2$M$_2$ PSII supercomplex with protein subunits labeled (PDB: 5XNL)\cite{20-su2017structure}. The solid line marks the separation of the two monomers.}
    \label{fig:enter-label}
\end{figure}

\section{Methods}
\subsection{Rate Matrix Construction}
We employ a structure-based quantum dynamical rate matrix $K$, where element $k_{ij}$ describes the energy transfer rate from exciton state $j$ to state $i$ in units of [ps$^{-1}$]. The rate matrix for the C$_2$S$_2$M$_2$ PSII supercomplex was constructed from the 5XNL structure of \emph{Pisum sativum}\cite{20-su2017structure} (Figure 1) and the C$_2$S$_2$ rate matrix was constructed from the 3JCU structure of \emph{Spinacia oleracea}\cite{19-wei2016structure} (Figure S2), as well as from available Hamiltonians for the protein subunits\cite{27-raszewski2008light,28-raszewski2005theory,29-muh2012structure,30-mascoli2020design,31-novoderezhkin2011intra}. The rates were quantified by organizing the Chl molecules into domains based on coulombic coupling strength and degree of exciton delocalization, then applying Modified Redfield Theory for intra-domain transfer and Generalized Forster Theory for inter-domain transfer\cite{16-bennett2013structure,32-yang2002influence} (see SI). Based on the inhomogeneous broadening widths reported in the literature, a random number is added to the site energy of each chlorophyll and 500 sets of site energies are generated to construct 500 rate matrices. The calculation of entropy is based on the averaged rate matrix from these 500 realizations. The rates satisfy detailed balance with respect to a Boltzmann distribution of each excitonic state, which allows for its thermodynamic interpretation\cite{33-limmer2024statistical}. Unless otherwise specified, calculations are performed with the C$_2$S$_2$M$_2$ rate matrix at 300K. 

\subsection{Free Energy Calculations}
Following an excitation, where the initial condition is defined as a single exciton or linear combination of sites in the PSII supercomplex, we evolve the population vector, $P(t)=\{p_1(t),p_2(t),\dots \}$, using the direct integration of the master equation,
\begin{equation}
P(t) = e^{K t}P(0)
\end{equation}
where $p_i(t)$ is the probability of exciton state $i$ being excited at time $t$. Entropy and enthalpy are calculated as $\Delta S = -k_B \sum_i p_i(t) \ln p_i(t)$ and $\Delta H = \sum_i p_i(t) E_i$ where $E_i$ represents the energy level of state $i$. The entropic and enthalpic components of the free energy are computed following,
\begin{equation}
\Delta G(t) =\Delta H (t) -T \Delta  S(t)		
\end{equation}
where we define $\Delta H(t)=\langle H(t)\rangle -H(0) $ as the enthalpy change and $\Delta S(t)=\langle S(t) \rangle -S(0)$ as the entropy change, relative to their initial conditions. Because we are working in the exciton basis, we define an excitation of Chl \textit{a} 509 in CP43, for example, as the exciton state that is most localized in that chlorophyll. We model open reaction centers by adding an irreversible trap in the RC of each dimer using phenomenological rates. We define the location of this trap as the midpoint between the reaction center PheoD1 (pheophytin) and ChlD1 molecules, which are predicted to form the charge transfer state\cite{34-nguyen2023charge} that we set as the lowest-energy state in the RC. To compute the distance of pigments from the open RC, we use the three-dimensional coordinates, determined from the crystal structure, of the central Mg atom for chlorophyll and the approximate center of the porphyrin ring for pheophytin. 

The enthalpy change describes both the scale and directionality (uphill or downhill) of the ensemble-averaged energy change of an exciton at specific times in the dynamics resulting from a defined initial condition. If the enthalpy maximum occurs after time zero, this point reveals the time before which the average population transfers against an energy gradient and after which the ensemble follows downhill excitation energy transfer (EET) to the RC trap. If the enthalpy change is negative, then the dynamics are driven by the availability of downhill energy transfer steps. 

\begin{figure*}[t]
    \centering
    \includegraphics[width=17cm]{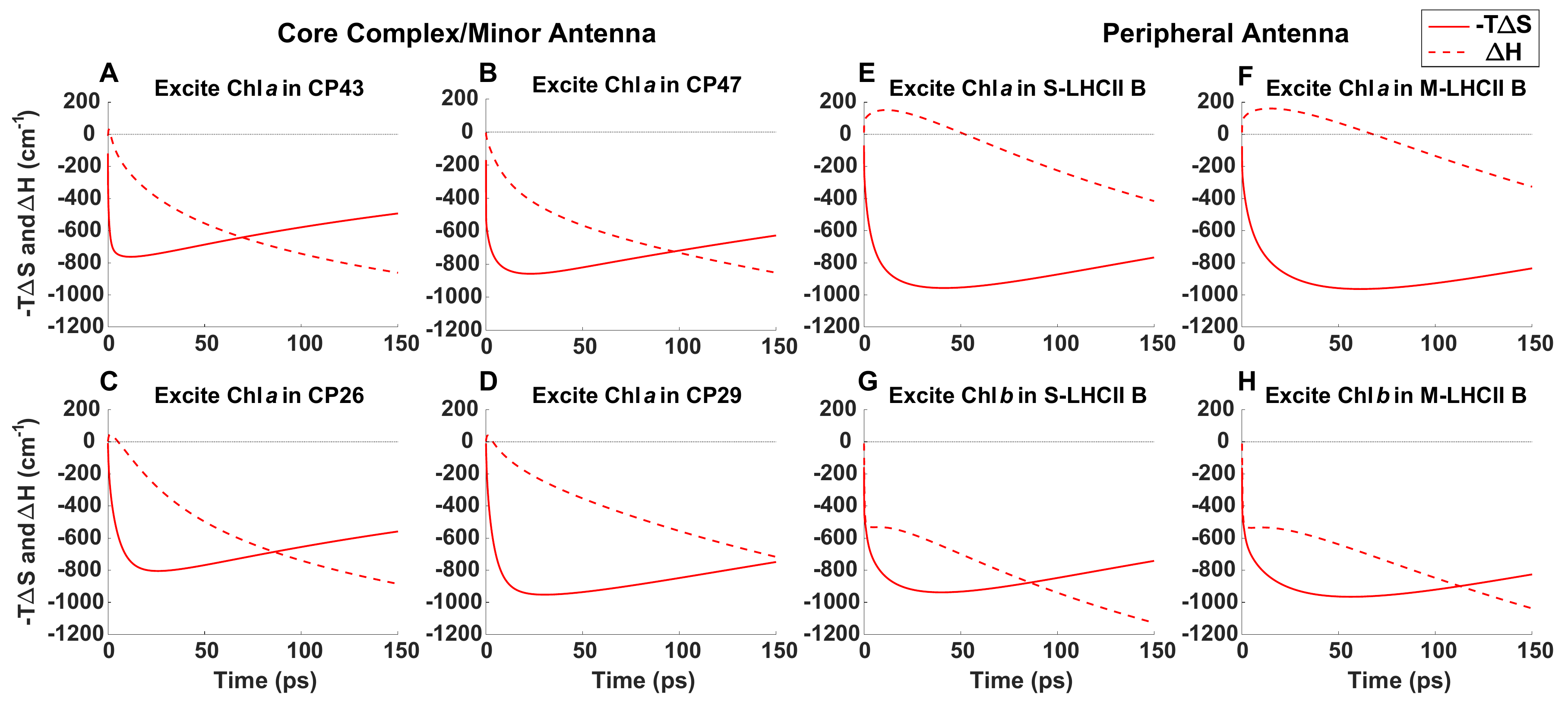}
    \caption{Entropy (solid line) and enthalpy (dashed line) components of the free energy change in time for initial excitations at selected states in the core antenna (A-B), minor antenna (C-D), and peripheral antenna (E-H) complexes. In order of A-E, excitations are localized in Chl \textit{a} 509 in CP43, Chl \textit{a} 610 in CP47, Chl \textit{a} 604 in CP26, Chl \textit{a} 604 in CP29, Chl \textit{a} 610 in S-LHCII (B), Chl \textit{a} 610 in M-LHCII (B), Chl \textit{b} 609 in S-LHCII (B), and Chl \textit{b} 609 in M-LHCII (B).}
    \label{fig:enter-label}
\end{figure*}
The entropy change, on the other hand, characterizes the spatial distribution of the ensemble EET trajectories at a given time. If the entropy increases, then ensemble energy transfer is driven by a high accessibility of nearby states. While the entropy is increasing, the ensemble population continues to spread out, reflective of individual trajectories following more divergent EET pathways. Likewise, when the entropy decreases, the ensemble of excitations becomes more directed and follows increasingly common pathways. Therefore, the time when the entropy is at a maximum is referred to as the population contraction time.

\section{Results and Discussion}
The entropic and enthalpic components of the free energy change are plotted in Figure 2. As entropy increases, the plotted entropy component, $-T\Delta S(t)$, decreases, so the minimum point on the plot corresponds to the maximum entropy or the population contraction time. Following excitation, the entropy for each initial condition rapidly increases, indicating population dispersion among nearby exciton states on sub-picosecond (ps) timescales. For the core excitations localized in Chl \textit{a} 509 in CP43 and Chl \textit{a} 610 in CP47, the ensemble population contracts at 12 ps and 23 ps, respectively (Figure 2). For excitations in the minor antenna in Chl \textit{a} 604 in CP26 and Chl \textit{a} 604 in CP29, the ensemble population contracts at 26 ps and 30 ps, respectively. S-LHCII (B) in the peripheral antenna exhibits population contraction times of 41 ps and 40 ps for respective Chl \textit{a} 610 and Chl \textit{b} 609 excitations, while similar excitations in M-LHCII (B) result in 61 ps and 56 ps contraction times. 

These times provide insight into the propensity of the initial condition to direct excitations directly toward the RC for trapping or to first spread the excitations out among the pigments. Initial excitations with slower ensemble population contraction toward the RC, and, likewise, a larger contribution from entropy (Figure S3), are designed to be well connected to many other states. Likewise, states with faster ensemble population contraction times are connected to energetically downhill transfer pathways for more rapid transfer toward the RC. 

The enthalpy displays highly variable behavior following excitation in different initial conditions. In some cases, the ensemble population transfers uphill in energy, resulting in an increase in the average enthalpy. For initial excitations in Chl \textit{a} 509 in CP43 (Figure 2A) and Chl \textit{a} 610 in CP47 (Figure 2B), this uphill energy transfer is minimal, with the enthalpy maximum occurring within 0.50 ps of initial excitation. Excitations in the minor antenna (Figure 2C and D) result in slightly longer-lived uphill EET until 1.30 ps and 0.95 ps for respective excitations in Chl \textit{a} 604 in CP26 and Chl \textit{a} 604 in CP29. For excitations localized in Chl \textit{a}, which account for 214 of the 316 excitations, the entropy term changes much more rapidly than the enthalpy term at early times. 

Peripheral Chl \textit{a} excitations (Figure 2E and F) result in much longer-lived uphill population transfer. The enthalpy terms for Chl \textit{a} 610 excitations in S-LHCII (B) and M-LHCII (B) maximize at 12 ps and 15 ps, respectively. The maximum enthalpy value, however, is less than 200 cm$^{-1}$, or typical thermal energies at room temperature, so this long-lived enthalpy increase is more representative of gradual, rather than energetically steep, uphill energy transfer steps. The states with this long-lived uphill energy transfer tend to be among the lowest-energy states in the complex, but this relationship is not perfectly correlated (Figure S4). Initial excitations in Chls \textit{a} 610 and 612 of the LHCII (B) complexes display the longest-lasting uphill energy transfer and likewise the longest excitation retention. States with long population retention should provide good candidates for nonphotochemical quenching (NPQ) sites and Chls \textit{a} 610 and 612 in LHCII have previously been suggested to be the sites of NPQ\cite{35-ruban2022chlorophyll,36-grondelle2006energy,37-ruban2007identification}. These relatively long-lasting ensemble EET dynamics against an energy gradient also highlight that entropically-controlled dynamics do not have to be confined to very short times. 

\begin{figure}
    \centering
    \includegraphics[width=8.5cm]{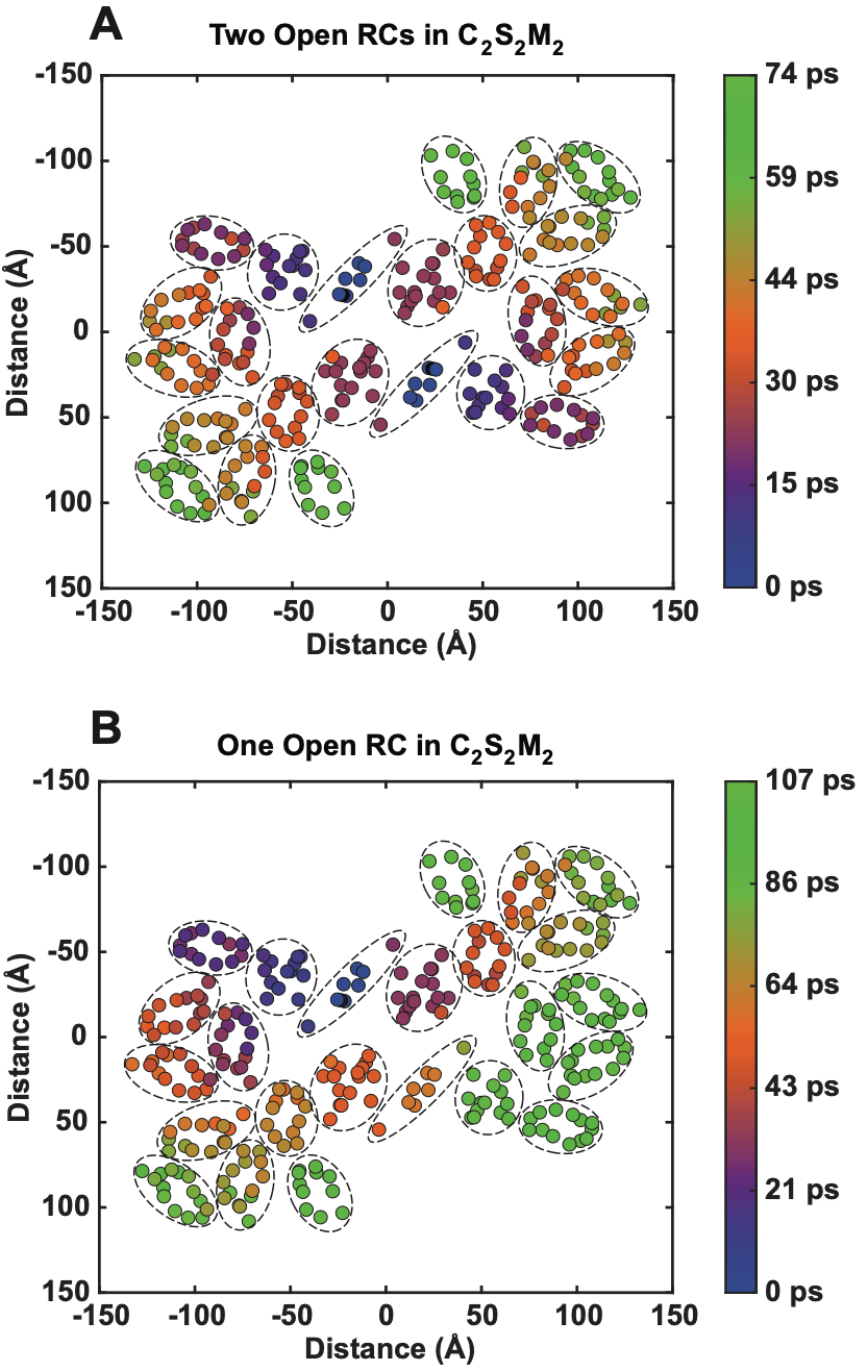}
    \caption{Population contraction time for each initial excitation projected onto the site basis for PSII when (A) both RCs are open and (B) the monomer 2 (bottom) RC is closed. The charge transfer state is located between ChlD1 and PheoD1 in the reaction center.}
    \label{fig:enter-label}
\end{figure}

Excitations localized in Chl \textit{b} in the peripheral antenna (Figure 2G and H), in contrast, result in exclusively downhill energy transfer in two kinetic phases: first, a rapid (sub-ps) phase followed by a slower (sub-ns) phase toward the RC. The fast timescale likely represents intra-complex equilibration, while the slower component represents inter-complex EET starting around 1 ps. When both reaction centers in the PSII dimer are open, the population contraction time for all initial excitations is less than 75 ps, which is about half of the complex’s 159 ps fluorescence lifetime, simulated from an initial excitation spread among all Chls \textit{a} in PSII (Figure 3). There is a general pattern of later population contraction times for states that are more peripheral to the complex, but this is not a very strongly correlated relationship with an R2 value of 0.68 (Figure S5). Therefore, the distance of a state from the RC is not sufficient to explain its population contraction time.

Photosystem II is a symmetric homodimer with two reaction centers located near the center of the complex and multiple protein subunits holding a fixed number of chlorophyll molecules in consistent orientations\cite{20-su2017structure}. When the population contraction times are compared across all PSII excitations, consistencies arise among states in the same protein subunit, particularly the core and minor antenna subunits (Figure S6). Excitations in the core CP43 subunit next to the D1 protein in the RC have an average population contraction time of 13 ps and a very high (81\%) average probability of being trapped in the same-monomer reaction center (Figure 4, Figure S7). Excitations in the minor antenna CP26 subunit on the same side of the RC, termed the D1 side, have an average population contraction time of 22 ps and a similarly high (80\%) average probability of same-monomer RC trapping. In contrast, excitations in the core CP47 subunit next to the D2 protein in the RC have a 24 ps average population contraction time and a 59\% average probability of same-monomer trapping. Excitations in the minor antenna CP29 subunit on the same side of the RC, termed the D2 side, have a 33 ps average population contraction time and a 51\% average probability of same-monomer trapping.  The faster population contraction times and higher probability of same-monomer trapping on the D1 side subunits (CP43 and CP26) is consistent with experimental and computational observation of faster EET and trapping on the D1 side\cite{13-leonardo2024bidirectional,27-raszewski2008light,38-yang2022ultrafast,39-deweerd2002pathways}.

\begin{figure}[t]
    \centering
    \includegraphics[width=8.5cm]{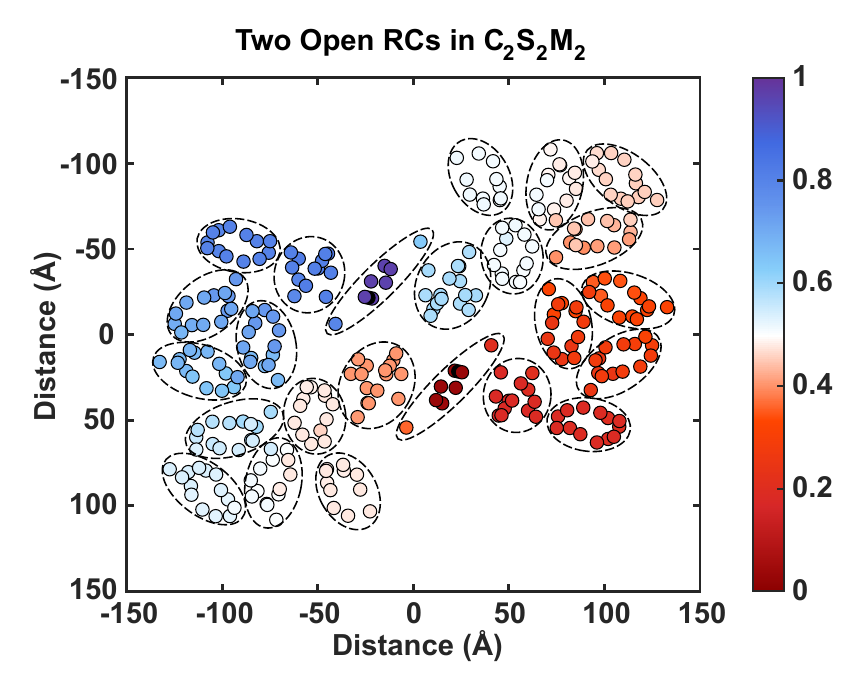}
    \caption{Probability that an initial excitation will be trapped in the monomer 1 reaction center (top), projected onto the site basis. }
    \label{fig:enter-label}
\end{figure}
The population contraction times among excitations in the S- and M-LHCII trimers are less similar than in the core and minor antenna subunits (Figure 4, Figure S7). For S-LHCII, subunit C, which is closest to CP43 (Figure 1), has the shortest average contraction time at 28 ps. S-LHCII subunits A and B, meanwhile, display 39 ps and 43 ps respective contraction times. All three S-LHCII subunits (A, B, and C) display high (72\%, 66\%, and 71\%) probability of trapping in the same-monomer RC, consistent with other D1-side subunits. For M-LHCII, subunits A and C display similar population contraction times of 45 ps and 47 ps, while subunit B, which is most peripheral (Figure 1), is slower at 61 ps. All three M-LHCII subunits (A, B, and C) display minimal (46\%, 49\%, and 51\%) preference for trapping in either monomer RC, consistent with other D2-side subunits. 

The faster population contraction times and higher probability of same-monomer trapping for D1-side subunits (CP43, CP26, and S-LHCII) suggest they are designed to achieve efficient and reliable trapping of excitations. The slower population contraction times and near-agnostic preference for either RC of D2 subunits (CP47, CP29, and M-LHCII) suggest they are designed to perform bidirectional energy transfer both toward and away from the same-monomer RC to facilitate inter-monomer energy transfer. 

The high similarity in the timescale of ensemble population dispersion and contraction among excitations within some protein subunits likely results from rapid intra-subunit equilibration, so inter-protein EET dynamics are indifferent to the specific state in the subunit that was excited. The core complex (CP43 and CP47) and minor antenna (CP26, CP29, and CP24) subunits have very similar population contraction times, with standard deviations up to ±4 ps. This suggests that these proteins act as distinct units in the PSII energy transfer network. Characterizing each protein’s role may help explain the design utility of PSII’s distinct subunits, as opposed to a more homogeneous photosystem design like photosystem I\cite{40-gobets2001energy} or bacterial phycobilisomes\cite{41-sohoni2023phycobilisome}.

Less similar times of population dispersion and contraction for excitations within the same protein subunit may result from rates of inter-protein EET that are comparable to or greater than rates of intra-protein equilibration. The diversity of free energy dynamics among individual states in the LHCII subunits, whose population contraction times have standard deviations up to $\pm$8 ps, suggests that this is precisely the case for states in the peripheral antenna. 

To further investigate the design of PSII, we consider the free energy terms for another common condition of the system. Natural light levels frequently fluctuate and the reaction centers of PSII intermittently close as a result of a recent charge separation\cite{1-blankenship2022molecular,42-amarnath2016multiscale,43-sipka2021light,44-slattery2018impacts}. Figure 3B demonstrates the population contraction times for each excitation when the reaction center in monomer 2 (bottom) is closed. The symmetry across the two monomers observed with two open RCs (Figure 3A) is now broken as excitations cannot be quenched until they encounter the open RC in monomer 1. Even with one closed RC, the slowest population contraction time in the complex is 107 ps, which is less than one-third of the 349 ps computed fluorescence lifetime for a closed monomer 2 RC (Figure 3B). This means that excitations across both PSII monomers are directed toward the open RC on timescales faster than the fluorescence decay rate, ensuring efficient charge separation is maintained even with one RC closed. 

The slowest population contraction times are not observed for the states in the now-closed monomer 2 RC, but rather for states in the D1-side subunits of that RC. The states within the closed reaction center have population contraction times around 62 ps (Figure S8). Excitations in CP47 and CP29 on the D2 side of the closed RC have population contraction times within 50 ps and 65 ps, respectively. In contrast, excitations in CP43 and CP26 on the D1 side of the closed RC have respective population contraction times around 90 ps and 100 ps. Excitations in the S-LCHII subunits in monomer 2 have the slowest population contraction times up to 107 ps when the monomer 2 RC is closed. Meanwhile, the free energy dynamics in the M-LHCII complex on the D2 side of the closed RC are minimally impacted. Evidently, excitations in the D1-side subunits are the most impacted by the closure of the corresponding RC. 

The disproportionate elongation of population contraction times for D1-side subunits compared to D2-side subunits of a closed RC further supports the notion that D1-side subunits are designed to perform reliable energy trapping by directing excitations to the reaction center more quickly. The resilience of D2-side subunit excitations in maintaining relatively fast population contraction times when an RC closes supports the notion that the D2-side subunits are designed to provide many alternate energy transfer pathways for excitations to reach a trap through entropically-dominated ensemble dynamics.

\begin{figure}[t]
    \centering
    \includegraphics[width=8.5cm]{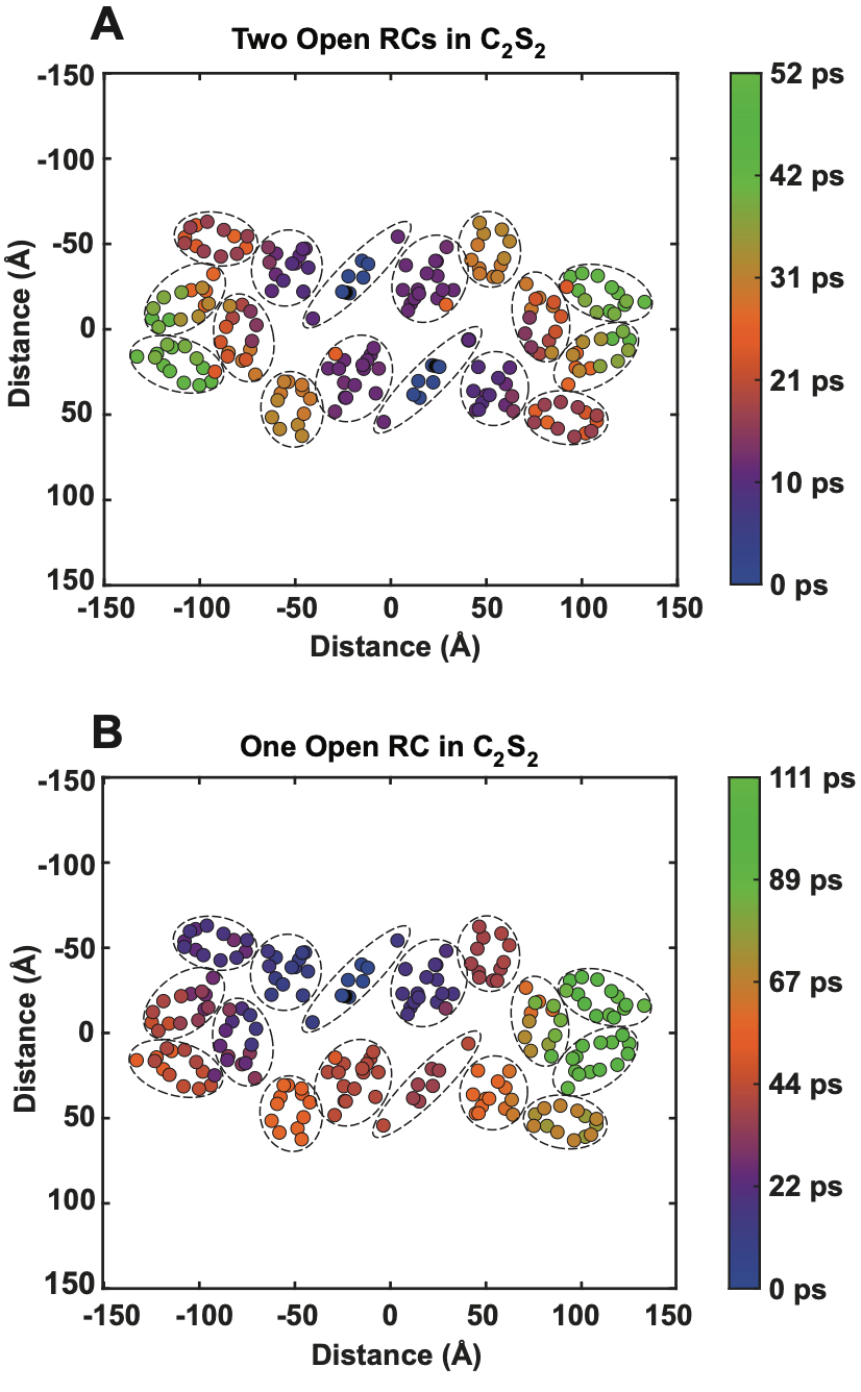}
    \caption{Population contraction times for the C$_2$S$_2$-type PSII supercomplex with (A) two open RCs and (B) the monomer 2 RC closed, projected onto the site basis. }
    \label{fig:enter-label}
\end{figure}
Under light stress conditions, the M-LHCII subunits may be removed\cite{45-horton2012optimization,46-albanese2016dynamic,47-kouril2013high,48-croce2020beyond}, forming the C$_2$S$_2$-type PSII supercomplex, which is the most common form in high light in Arabidopsis thaliana\cite{46-albanese2016dynamic}. Figure 5 shows the population contraction times for the C$_2$S$_2$M$_2$ supercomplex. With both RCs open, the longest population contraction time is 52 ps, compared to 74 ps for the C$_2$S$_2$M$_2$ supercomplex. When one reaction center closes, the population contraction times become slightly slower in the C$_2$S$_2$ complex, with a 111 ps maximum, than in the larger C$_2$S$_2$M$_2$ complex, with a 107 ps maximum. 

For both supercomplexes, the longest population contraction times for the closed monomer-2 RC case occur in the S-LHCII subunit. Returning to our example Chl \textit{a} 610 excitation in S-LHCII (B), we observe population contraction times of 41 ps in C$_2$S$_2$M$_2$ and 42 ps in C$_2$S$_2$ when both RCs are open (Table S1). When the monomer-2 RC is closed and the same S-LHCII state on the monomer 2 side is excited, these times become 87 ps in C$_2$S$_2$M$_2$ (Figure S9) and 101 ps in C$_2$S$_2$, corresponding to times that are respectively 2.1x and 2.4x as long as the two open RCs case. This phenomenon of disproportionate lengthening of population contraction times in C$_2$S$_2$ upon an RC closure is also observed for core and minor antenna subunits on the D2 side. For our example Chl \textit{a} 610 excitation in CP47, the ensemble population contracts at 23 ps in C$_2$S$_2$M$_2$ and 12 ps in C$_2$S$_2$ when both RCs are open (Table S1). However, when the monomer 2 RC closes and the CP47 state in monomer 2 is excited, these times become 49 ps for C$_2$S$_2$M$_2$ and 42 ps for C$_2$S$_2$, which are respectively 2x and 3.5x as long as the two open RCs case. 

This same pattern is not, however, present for CP43 excitations on the D1 side. Closing the monomer 2 RC results in a Chl \textit{a} 509 excitation in the monomer-2 CP43 subunit displaying population contraction times about 8x and 6x as long as the two open RC case for respective C$_2$S$_2$M$_2$ and C$_2$S$_2$ excitations. 

If the only role of the M-LHCII complexes is to increase the absorption cross-section for the supercomplex, then the longer population contraction times for C$_2$S$_2$ than C$_2$S$_2$M$_2$ following an S-LHCII excitation or a core or minor antenna excitation on the D2 side is an unexpected result, since C$_2$S$_2$M$_2$ clearly has more states for excitations to explore. This result indicates that the M-LHCII complexes play an important role in facilitating inter-monomer EET, in line with observations by Yang et al\cite{12-yang2024design}. Not only is inter-monomer EET useful to maintain efficient light harvesting in stress conditions, demonstrated by the D2-side subunits when one RC is closed, but it can also serve to increase the efficiency of light harvesting under typical conditions by providing peripheral excitations multiple options to reach a quenching site. This demonstrates the capacity of M-LHCII subunits to create numerous, and potentially more efficient, energy transfer pathways to an open RC that are not present in C$_2$S$_2$. 

Clearly, the ability to add and remove LHCII complexes allows PSII to regulate its free energy network, especially the entropy component, by adjusting the number of states that different complexes are connected to\cite{18-croce2014natural,48-croce2020beyond,49-caffarri2011excitation}. Photosynthetic organisms may control the presence of LHCII, and therefore the entropic character of the PSII free energy landscape, to optimize the energy transfer network to their local environment. This work demonstrates the significance of both free energy terms and the ability to regulate them in the design of energy transfer networks.

\section{Concluding Remarks}
rIn an attempt to uncover the design principles underlying the PSII energy transfer network, we studied the free energetics accompanying relaxation of excitations in each state in the PSII supercomplex. The ensemble level calculations described above complement the individual trajectory calculations of Yang et al\cite{12-yang2024design} and bring a different perspective to the design of the Photosystem II supercomplex. The current calculations show that, at the ensemble level, there are two phases to motion of excitons: an initial entropy-driven phase in which the ensemble explores the landscape and a second enthalpy-driven phase of more directed motion toward the reaction center. The first phase can last for up to 50\% of the average exciton lifetime, but this component can be modified by the PSII supercomplex via addition or removal of LHCII subunits. Clearly, entropy is being used as a design principle to optimize the structure of the PSII energy transfer network as well as to dynamically respond to different environmental conditions.

We found that the PSII supercomplex contains three primary classes of energy transfer patterns. Some excitations, primarily in D1-side subunits, are well connected to downhill EET steps and display directed transfer toward the trap. Other excitations, particularly in CP47 and CP29 on the D2 side, spread energy out bidirectionally to the RC and periphery. Excitations in the LHCII subunits, particularly M-LHCII, spread energy between the two monomers. Working together, these pathways ensure (1) efficient energy trapping in the RC, (2) effective NPQ in the periphery, and (3) robust energy trapping in the other-monomer RC if the same-monomer RC is closed. 

We would not expect to observe these unique features of PSII if its energy landscape were structured as a funnel or as a homogeneous landscape. The variations in connectivity of individual protein subunits makes these different EET patterns possible, demonstrating that employing multiple subunits can provide a selective advantage and even increase quantum efficiency of energy transfer. PSII’s dual goals of energy trapping and photoprotection are realized through the varied structure of the free energy network across the complex. States in subunits designed to accomplish reliable energy trapping are engineered to be well-connected (have fast transfer) to states with lower energy levels in the direction of the reaction center. To maintain resilience of the energy transfer network, alternate options must be available for trapping, hence the dimeric design of PSII. States designed to maintain PSII energy transfer in varying conditions must be well-connected to many other states, thereby allowing entropy to dominate ensemble dynamics and create diverse energy transfer pathways. Under low-light conditions, PSII can add M-LHCII subunits, which not only expand the light harvesting capacity of the complex, but also increase the connectivity of the two monomers. This illustrates another entropy-motivated design principle in PSII, whereby the complex increases the density of states between the two monomers to promote inter-monomer transfer and increase light harvesting rates for certain excitations. As PSII adds and removes LHCII complexes, it is dynamically tuning its entropic landscape, which has significant implications for the energy transfer network. This demonstrates the capacity for entropy to act as a tunable design principle in complex energy transfer systems, and illustrates its significance in accomplishing diverse and variable energy transfer goals. 

\section*{Acknowledgments}
This research was supported by the U.S. Department of Energy, Office of Science, Basic Energy Sciences, Chemical Sciences, Geosciences, and Biosciences Division. J.L.H. and S.-J.Y acknowledge financial support from the Kavli Energy Nanoscience Institute at Berkeley. D. T. L is supported by the U.S. Department of Energy, Office of Science, Basic Energy Sciences, CPIMS Program Early Career Research Program under Award DEFOA0002019 and the Alfred P. Sloan Foundation. All authors thank Doran Raccah for code availability.

\section*{Contributions}
J.L.H. proposed the project and performed all calculations with assistance from S.J.Y. All authors contributed to interpretation of results. J.L.H and G.R.F. drafted the manuscript and all authors contributed to manuscript revisions.

\section*{Contact}
johannahall@berkeley.edu (J.L.H.); yangsj@mit.edu (S.-J.Y.); dlimmer@berkeley.edu (D.T.L.);  grfleming@lbl.gov (G.R.F.)

\section*{Competing Interests Statement}
The authors declare no competing interests.

\section*{Data availability}
The code for constructing the rate matrix is available at https://zenodo.org/records/13346121. The authors declare that all study data are included in the article and/or supporting information. The data presented in this study are available from the corresponding authors upon reasonable request.

\bibliography{ref.bib}
\pagebreak

\end{document}